\documentclass[acmsmall]{acmart}

\usepackage{etex}
\usepackage{subcaption} %% For complex figures with subfigures/subcaptions
\let\terms\undefined

\usepackage{times}            % standard fixed width font
\usepackage{graphicx}
\usepackage{amsmath}
\usepackage{amssymb}

\usepackage{xspace}
\usepackage{footnote}
\usepackage{amsfonts}
\usepackage{hhline}
\usepackage{multirow}
\usepackage{setspace} 
\usepackage{epsfig}
\usepackage{url}
\usepackage{booktabs}
\usepackage{xcolor}
\usepackage{lipsum}
\usepackage{courier}
\usepackage{listings}
\usepackage{caption}
\usepackage{colortbl}
\usepackage{arydshln} %dashed lines
\usepackage{makecell} %format table cells
\usepackage{tikz}
\usepackage{varwidth}
\usetikzlibrary{fit,arrows,calc,positioning,quotes}

\usepackage{textcomp}

\tikzset{
  -|-/.style={
    to path={
      (\tikztostart) -| ($(\tikztostart)!#1!(\tikztotarget)$) |- (\tikztotarget)
      \tikztonodes
    }
  },
  -|-/.default=0.5,
  |-|/.style={
    to path={
      (\tikztostart) |- ($(\tikztostart)!#1!(\tikztotarget)$) -| (\tikztotarget)
      \tikztonodes
    }
  },
  |-|/.default=0.5
}

\long\def\com#1{}
\newcommand{\app}{VeriCI\xspace}

\newcommand{\repoCode}[1]{\ensuremath{R_{#1}}\xspace}
\newcommand{\repoStatus}[1]{\ensuremath{S(R_{#1})}\xspace}
\newcommand{\repoSummary}[1]{\ensuremath{\hat{R}_{#1}}\xspace}
\newcommand{\repoModel}[1]{\ensuremath{M(\hat{R}_{#1})}\xspace}

\newcommand{\trainingSet}{\ensuremath{\mathcal{TR}}\xspace}
\newcommand{\reals}{\ensuremath{\mathbb{R}}}

\renewcommand{\implies}{\ensuremath{\Rightarrow}}

\newcommand{\para}[1]{\smallskip\noindent {\bf #1}}

\newcommand{\ie}{{\em i.e.\xspace}}
\newcommand{\eg}{{\em e.g.\xspace}}

\newcommand{\squishlist}{
   \begin{list}{$\bullet$}
    { \setlength{\itemsep}{0pt}      \setlength{\parsep}{3pt}
      \setlength{\topsep}{3pt}       \setlength{\partopsep}{0pt}
      \setlength{\leftmargin}{3.5mm} \setlength{\labelwidth}{1em}
      \setlength{\labelsep}{0.5em} } 
}

\newcommand{\squishend}{
    \end{list}  }

\makeatletter\if@ACM@journal\makeatother
\setcopyright{acmcopyright}
\acmPrice{}
\acmYear{2018}
\copyrightyear{2018}
\acmVolume{1}
\begin{document}

%% Title information
\title{Statically Verifying Continuous Integration Configurations}
                                        %% [Short Title] is optional;
                                        %% when present, will be used in
                                        %% header instead of Full Title.
%\titlenote{with title note}             %% \titlenote is optional;
                                        %% can be repeated if necessary;
                                        %% contents suppressed with 'anonymous'
%\subtitle{Subtitle}                     %% \subtitle is optional
%\subtitlenote{with subtitle note}       %% \subtitlenote is optional;
                                        %% can be repeated if necessary;
                                        %% contents suppressed with 'anonymous'

%% Author information
%% Contents and number of authors suppressed with 'anonymous'.
%% Each author should be introduced by \author, followed by
%% \authornote (optional), \orcid (optional), \affiliation, and
%% \email.
%% An author may have multiple affiliations and/or emails; repeat the
%% appropriate command.
%% Many elements are not rendered, but should be provided for metadata
%% extraction tools.

%% Author with single affiliation.
\author{Mark Santolucito}
\orcid{0000-0001-8646-4364}             %% \orcid is optional
\affiliation{
  \department{Computer Science}              %% \department is recommended
  \institution{Yale University}            %% \institution is required
  \streetaddress{51 Prospsect St.}
  \city{New Haven}
  \state{CT}
  \postcode{06511}
  \country{USA}
}
\email{mark.santolucito@yale.edu}          %% \email is recommended
\author{Jialu Zhang}
\affiliation{
  \department{Computer Science}              %% \department is recommended
  \institution{Yale University}            %% \institution is required
  \streetaddress{51 Prospsect St.}
  \city{New Haven}
  \state{CT}
  \postcode{06511}
  \country{USA}
}
%\email{mark.santolucito@yale.edu}          %% \email is recommended

\author{Ennan Zhai}
\affiliation{
  \department{Computer Science}              %% \department is recommended
  \institution{Yale University}            %% \institution is required
  \streetaddress{51 Prospsect St.}
  \city{New Haven}
  \state{CT}
  \postcode{06511}
  \country{USA}
}
\email{ennan.zhai@yale.edu}          %% \email is recommended

\author{Ruzica Piskac}
\affiliation{
  \department{Computer Science}              %% \department is recommended
  \institution{Yale University}            %% \institution is required
  \streetaddress{51 Prospsect St.}
  \city{New Haven}
  \state{CT}
  \postcode{06511}
  \country{USA}
}
\email{ruzica.piskac@yale.edu}          %% \email is recommended

\iffalse
\titlenote{This research was sponsored by the NSF under grant
CCF-XXXXXXX}

\thanks{Author's addresses: Mark Santolucito, Jialu Zhange, Ennan Zhai, and 
Ruzica Piskac, Computer Science Department, Yale University}
\fi

%% Paper note
%% The \thanks command may be used to create a "paper note" ---
%% similar to a title note or an author note, but not explicitly
%% associated with a particular element.  It will appear immediately
%% above the permission/copyright statement.
%\thanks{with paper note}                %% \thanks is optional
                                        %% can be repeated if necesary
                                        %% contents suppressed with 'anonymous'

%% Abstract
%% Note: \begin{abstract}...\end{abstract} environment must come
%% before \maketitle command
\begin{abstract}
Continuous Integration (CI) testing is a popular software development technique that allows developers to easily
check that their code can build successfully and pass tests across various system environments. In order to use
a CI platform, a developer must include a set of configuration files to a code repository for specifying build
conditions. Incorrect configuration settings lead to CI build failures, which can take hours to run, wasting
valuable developer time and delaying product release dates. Debugging CI configurations is challenging because
users must manage configurations for the build across many system environments, to which they may not have
local access. Thus, the only way to check a CI configuration is to push a commit and wait for the build result.
To address this problem, we present the first approach, VeriCI, for statically checking for errors in a given
CI configuration before the developer pushes a commit to build on the CI server. Our key insight is that the
repositories in a CI environment contain lists of build histories which offer the time-aware repository build
status. Driven by this insight, we introduce the Misclassification Guided Abstraction Refinement (MiGAR) loop
that automates part of the learning process across the heterogeneous build environments in CI. We then use
decision tree learning to generate constraints on the CI configuration that must hold for a build to succeed by
training on a large history of continuous integration repository build results. We evaluate VeriCI on real-world
data from GitHub and find that we have 83\% accuracy of predicting a build
failure.

\end{abstract}

%% 2012 ACM Computing Classification System (CSS) concepts
%% Generate at 'http://dl.acm.org/ccs/ccs.cfm'.

%\ccsdesc[500]{Software and its engineering~General programming languages}
%\ccsdesc[300]{Social and professional topics~History of programming languages}
%% End of generated code

%% Keywords
%% comma separated list
  %% \keywords is optional

%% \maketitle
%% Note: \maketitle command must come after title commands, author
%% commands, abstract environment, Computing Classification System
%% environment and commands, and keywords command.
\maketitle

\section{Introduction}
\label{sec:intro}

%CI is so awesome and everyone is using it
Continuous Integration (CI) testing is seeing broad adoption with the increasing popularity of the GitHub pull-based development model~\cite{gousios2014exploratory}.
There are now a plethora of open-source, GitHub-compatible, cloud-based CI tools, such as Travis CI\cite{travisci}, CircleCI~\cite{circleci}, Jenkins~\cite{jenkinsci}, GitLab CI~\cite{gitlabci}, Codefresh~\cite{codefresh} and TeamCity~\cite{teamcity}.
Over 900,000 open source project are using TravisCI alone~\cite{travisci}.
In a typical CI platform, developers receive continuous feedback on every commit indicating if their code successfully built under the specified dependencies and environments and whether the code passed all the given test cases.
CI has been particularly effective in the popular decentralized, social coding contexts such as GitHub.
By automatically building and testing a project's code base, in
isolation, with each incoming code change (\ie, push commit, pull request), CI 
   speeds up development (code change throughput)~\cite{hilton2016usage, pham2013creating},
   helps to maintain code quality~\cite{vasilescu2015quality, gousios2016work}, and
   allows for a higher release frequency which ``radically decreases time-to-market''~\cite{leppanen2015highways}.
In one case, a product group at HP reduced development costs by 78\% by introducing a CI environment~\cite{hilton2016usage}.

% But CI has major problems
However, for all the benefits CI brings, it also suffers from some serious limitations.
CI builds for large projects can take hours, and in particularly painful cases, some builds can take almost 24 hours~\cite{sc-longbuild}.
These slow builds not only interrupt the development process, but if the CI build returns an error status it means the software could not even be built on the target environments and the tests were not able to run.
This means that the user cannot receive any feedback on the functional correctness of their code.
A recent usage study~\cite{travis-torrent} of TravisCI found that 15-20\% of failed TravisCI builds end in an error status.
Using the data from~\cite{travis-torrent}, we can observe that since the start of 2014, approximately 88,000 hours of server time was used on TravisCI projects that resulted in an error status.
This number not only represents lost server time, but also lost developer time, as developers must wait to verify that their work does not break the build.
As one example, the release of a new version of the open-source project, SuperCollider, was delayed for three days because of slow TravisCI builds~\cite{SC_Delay}

\begin{figure}[tbp]
  \centering
  \includegraphics[width=0.8\textwidth]{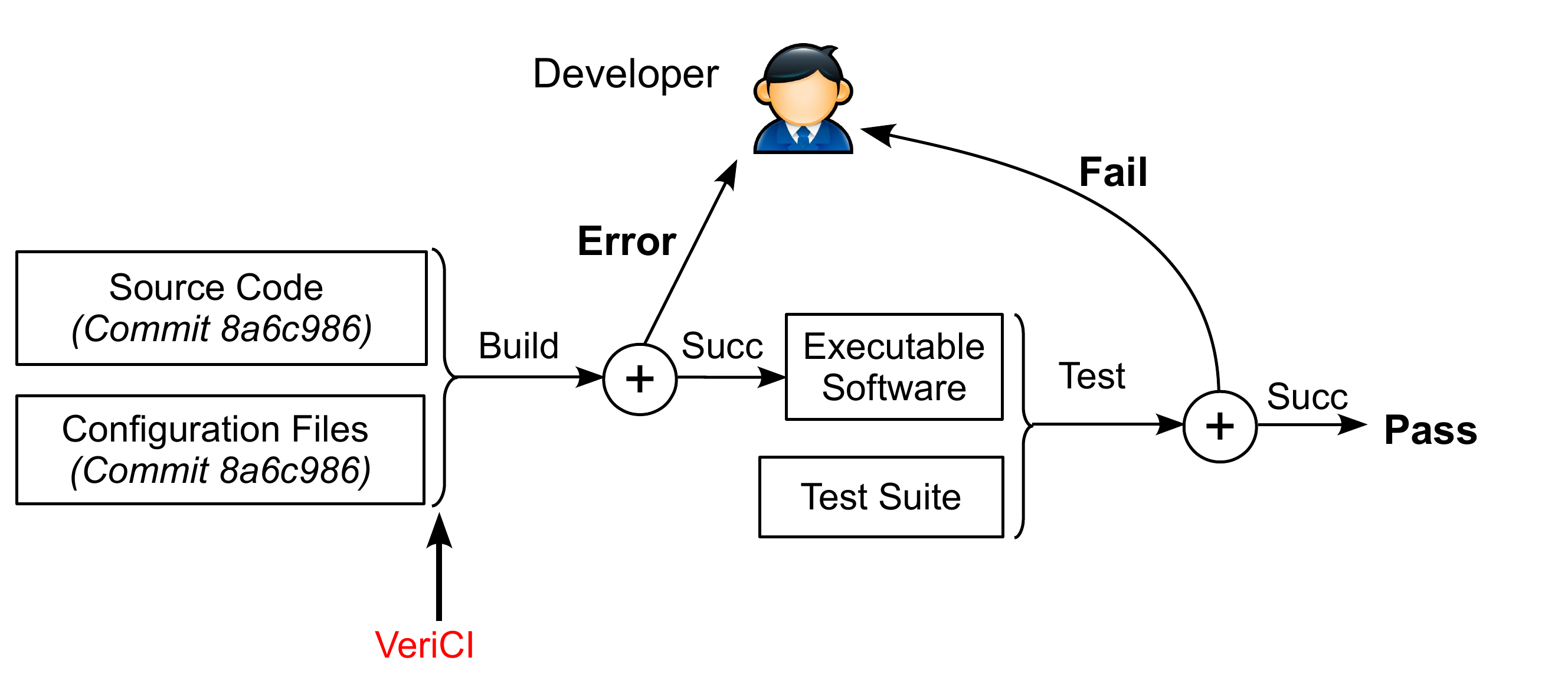}
  \caption{A typical build process of CI. \app is our tool that can statically analyze previous commits and predict it the current 
    will fail and if yes, what is the root cause.}
  \label{fig-build}
\end{figure}

% what is the critical source of these problems
%To understand the source of the erroring builds, 
Figure~\ref{fig-build} presents a typical CI build process.
A developer must include code and a test suite, and in addition, CI configuration files to the repository. These files specify build conditions such as the operating system, disk size, which compiler flags to use, which library dependencies are required, etc.
Then, each time a developer commits new code, the CI platform builds many versions of the executable software according to the configuration, runs this executable on the test suite with the appropriate hardware as operated by the CI provider, and returns the results.
If the executable was not built, the tests cannot be executed, and the developer must try to fix the configuration before any functional correctness can be checked.
The main sources of difficulty in debugging CI configurations 
include: 1) the developer does not know whether her configurations
can work across multi-platform, and 
2) the only way to check that is to
push a commit and wait for the CI provider to complete the build.
If these malformed configurations 
could be quickly and statically checked on the client side before
the build, the CI workflow could be considerably improved.

% what has been done about these problems, ending with why none of that will work
\para{Why the state of the art does not help?}
Building tools for configuration file support, management, and verification has been an active direction of research~\cite{xu16early,attariyan10automating,su07autobash,baset2017usable,raab2016elektra}.
Configuration files are so critical that domain specific language, like the industrial products Chef and Puppet, have been built for the express purpose of managing the complexity of configuration files.
Even with specialized tools and languages, there has been need for further work on the verification and repair of these configuration management languages~\cite{WeissGB17,ShambaughWG16}.
Other work has leveraged machine learning based methods to automatically generate specifications for configuration files~\cite{zhang14encore, SantolucitoOOPSLA17, wang04automatic}.
However, all the existing efforts have been focused on environments 
where the system semantics are known, 
such as database system setup, or network configuration 
for traffic forwarding.
In CI environments, the code is built on diverse platforms, 
and developers may be adding arbitrary new components or languages 
to the code base over time.
Furthermore, because CI configuration files evolve over time 
based on the newly committed programs, it is much more difficult to
detect the root causes of CI configuration errors
than the configuration errors in other systems, such as database system
and network configuration.
In order to handle the diverse use cases for CI and 
the dynamic ``structure'' of the configurations, 
a new approach specific to CI environment is highly needed.

\para{Our approach.}
To this end, we propose the first approach for 
statically checking the errors in a given CI configuration
before the developer builds this configuration on the CI server.
The main challenge with verifying such build configurations is that we
do not have a complete model for the semantics of the build process.
Such a model would need to account for a complex combination of factors,
such as heterogeneous languages, libraries, network states, etc.
Trying to build an exact model for 
the repository and its corresponding platforms is an prohibitively 
expensive engineering task.

Our key insight is that the repositories in CI environment contain 
lists of build histories which offer the 
time-aware repository status including source code, configuration
files, and build status (\ie, error, fail or pass).
Driven by the above insight,
we propose \app, a language agnostic approach that can derive
a set of system invariants to model configuration correctness of CI environments.

In \app, we leverage two unique properties of the CI environment.
First, we have a labelled dataset of build statuses over the commit history allowing us to use a supervised learning approach.
In particular we use decision tree learning, which has already seen application in the area of program verification~\cite{KrishnaPW15,sankaranarayanan2008dynamic,BrazdilCKT18}.
Second, we use the fact that in our training set, as a product of a commit history, each sample (commit) has an incremental change with respect to the previous sample (commit).
We use this property to create a semi-automated feature extraction procedure, allowing us to apply \app to a large scale training set of repositories from GitHub.
This feature extraction relies on the Misclassification Guided Abstraction Refinement (MiGAR) loop, which we base on the CEGAR loop~\cite{clarke2000counterexample}, 
Automated feature extraction for structured C-like languages has been explored in previous work~\cite{chae2017automatically}, but the highly heterogeneous structure of CI configurations presents new challenges.

Finally, \app discovers a model of configuration correctness for a repository that a user can use to statically predict if the build process will fail.
\app additionally uses this model to provide feedback to the user in the form of error messages so that the user might fix their build.
We ran \app on 33 large repositories from GitHub and correctly predicted the build status 83\% of the time. 
\app was able to provide a justification for its prediction, and 38\%-48\% of the time, the justification matched the error that the user actually used to fix their repository.

In summary, our work makes the following contributions.

\begin{itemize}

\item We give a formal language to describe the CI build process and associated terms that allows for clearer reasoning about CI environments.

\item We introduce the Misclassification Guided Abstraction Refinement (MiGAR) loop, for a semi-automated feature extraction in the context of a machine learning for verification.

\item We describe how to use decision tree learning to achieve explainable CI build prediction that provides justification for the predicted classification as an error message to the user.

\item We implement a tool, \app, for learning models of correctness of CI configurations based on the above methods. We evaluate our tool on real world data from GitHub and find we have a 83\% accuracy of predicting a build failure. Furthermore, \app provides an explanation for the classification as an error report to the user.

\end{itemize}

\section{Motivating Examples}
\label{sec-motivation}

In a continuous integration (CI) environment, build and testing times of large projects can be on the scale of hours.
When a CI build errors,  the tests cannot be run and developers do not have a complete picture of the status of the code.
As an additional burden on developers, if a build fails, they must spend time to find the cause of the failure and find an appropriate fix.
We give here examples of the kinds of failures that developers face in the CI environment, and show how our tool, \app, can help guide developers to the required fix.

\subsection{Identifying an error within a single file}
\label{sec:example1}

We demonstrate the functionality of \app by using a real-world example of a TravisCI build failure from the \texttt{sferik/rails\_admin} repository~\cite{sferik-break} on GitHub.
This is a large repository with (at the time of writing) 4342 commits over 347 contributors, 6766 stars, and 2006 forks.
We will show the difficulty in identifying the cause of failure, and also demonstrate how \app can warn the developer about a potential build failure.

\begin{figure}[h]
  \centering
  \includegraphics[width=0.99\textwidth]{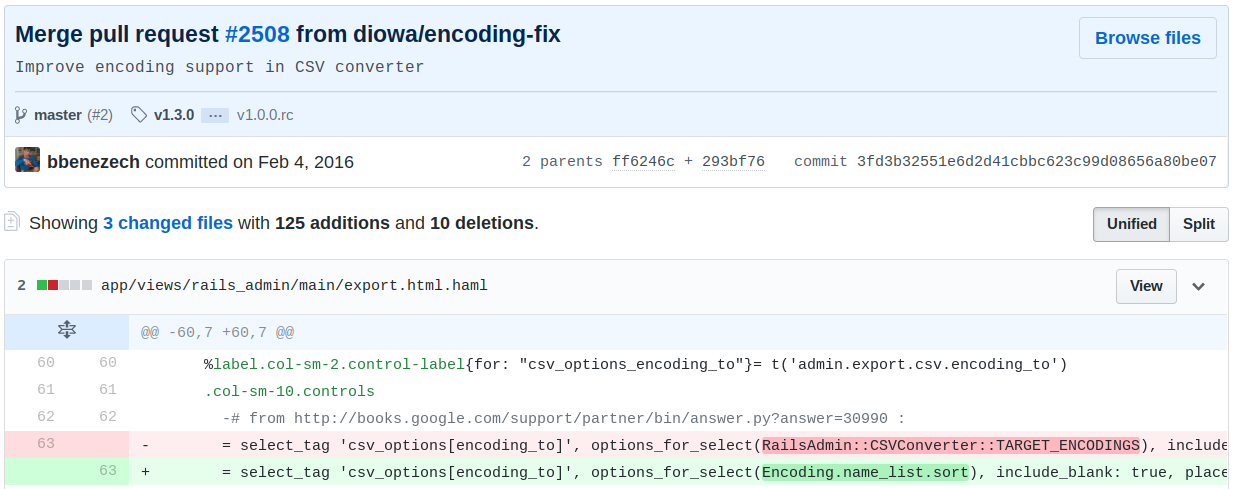}
  \caption{The commit that broke the build of \texttt{sferik/rails\_admin} and took 20 days to fix.}
  \label{fig-motivation}
\end{figure}

Fig.~\ref{fig-motivation} shows a commit to the \texttt{sferik/rails\_admin} repository, where an administrator merged a pull request (a set of changes) into the master branch, which caused the TravisCI build to error.
In the pull request, which changed 3 files by adding 125 lines and deleting 10 lines, the user added the message that this pull request should ``improve encoding support in CSV converter''.
However, after merging this pull request, TravisCI reported that it was not able to successfully build.
Manually checking the log information is a tedious process and it is very difficult to gain any helpful information for understanding or correcting the issue. If we inspect the next 14 commits in the repository, they still have the same failed TravisCI status. It was not until 20 days later that a contributor to the repository was able to correct this build error~\cite{sferik-fix}.
 
To test the functionality of \app, we run it on the pull request that caused the failure. 
\app reports that it (correctly) predicts a build failure.
In addition, it also provides an explanation for its classification, by saying that the two keywords  \texttt{TARGET\_ENCODINGS=\%w(UTF-} and \texttt{-rvm} are likely to be contributing to the build failure.
\app also points out to the user the locations in the code where these keywords appear so that the user can address this issue.

\begin{minipage}{\linewidth}
\begin{lstlisting}[language=C, xleftmargin=.01\textwidth]
Predicted build failure based on potential error locations:
lib/rails_admin/support/csv_converter.rb:Line 7 
   TARGET_ENCODINGS=%w(UTF-
.travis.yml:Multiple Lines
   -rvm
\end{lstlisting}
\end{minipage}

In fact, this is exactly the location of the change made by the user who fixed the build error~\cite{sferik-fix}.
The user fixed this change by reverting most of the changes from the breaking pull request, along with the comment ``Simplify export by trusting DB driver's encoding handling''.
While the user made a change that both fixes the build error and
simplifies the code, \app has identified the ``minimal'' problem, \ie, the smallest set of commands that the build system was not able to handle.
We envision that \app will not only be used to effectively predict a build failure, but also to point the user towards locations of potential issues. 
Since there may be many solutions to a build error, \app does not try to suggest a repair, but rather helps the user identify the root cause of a build failure. This way the users can address the problem in whichever way they think is most appropriate.

\subsection{Identifying errors spanning multiple files}
\label{sec:example2}

In previous example, \app detected the root cause of an error which required changing a single line.
However, many CI errors can be a result of complex relationships between multiple files and branches in a repository.
To demonstrate this, we take another real-world example, coming from the \texttt{activescaffold/active\_scaffold}~\cite{activescaffold-break} repository which has (at the time of writing) 4961 commits over 87 contributors, 955 stars, and 313 forks.

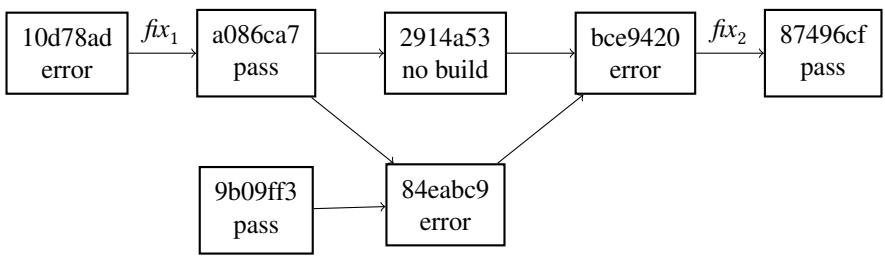
\begin{figure}[h!]
  \centering
  \begin{tikzpicture}
\tikzstyle{b} = [rectangle, draw, node distance=0.9cm, minimum width=1.5cm, minimum height=2em, thick, align=center, inner sep=0.2cm
                , execute at begin node={\begin{varwidth}{15em}}
                , execute at end node={\end{varwidth}}]
\tikzstyle{c} = [fill=gray!30, rounded corners=5pt]
\tikzstyle{l} = [draw, -latex',thick]

    \node [b] (10d7) {10d78ad \\ error};
    \node [b, right=of 10d7] (a086) {a086ca7 \\ pass};
    \node [b, below=of a086] (9b09) {9b09ff3 \\ pass};
    \node [b, right=of a086] (2941) {2914a53 \\ no build};
    \node [b, below=of 2941] (84ea) {84eabc9 \\ error};
    \node [b, right=of 2941] (bce9) {bce9420 \\ error};
    \node [b, right=of bce9] (8749) {87496cf \\ pass};

    \path[->] 
	(10d7) edge ["$\textit{fix}_1$"] (a086)
    	(a086) edge (84ea)
        (a086) edge (2941)
        (9b09) edge (84ea)
        (84ea) edge (bce9)
        (2941) edge (bce9)
        (bce9) edge ["$\textit{fix}_2$"] (8749);
\end{tikzpicture}
  \caption{A GitHub history chain depicting dependencies between commits}
  \label{fig:activescaffold-hist}
\end{figure}

Fig.~\ref{fig:activescaffold-hist} shows a sequence of commits. 
Each commit is represented by a box and the arrows denote the parent commit. 
A commit can have two child commits (as in commit a086ca7) when two users make a different change based on the same state of code. This difference was eventually resolved when a commit has two parent commits (as in the case of commit bce9420).

To better understand these commits, the {\tt{git diff}} command can show the difference between two different commits. 
Fig.~\ref{fig:activescaffold-code} depicts the \texttt{git diff} from a selection of commits from Fig.~\ref{fig:activescaffold-hist}. Italics indicate which file was changed, while (+/-) indicate what was added or removed.

{
\begin{figure}[!h]
    \centering
    \hspace{1cm}
    \begin{minipage}{.4\textwidth}
	\begin{lstlisting} [label={lst:repoCode},language=C,caption={10d78ad},mathescape=true]
$version.rb$
module Version
 MAJOR = 3
 MINOR = 4
- PATCH = 34
+ PATCH = 35
	\end{lstlisting}
    \end{minipage}
    \hspace{1cm}
    \begin{minipage}{0.4\textwidth}
	\begin{lstlisting} [label={lst:repoSummaryx},language=C,caption={a086ca7},mathescape=true]
$Gemfile.lock$
- active_scaffold (3.4.34)
+ active_scaffold (3.4.35)


$ $
        \end{lstlisting}
    \end{minipage}
    \\
    \hspace{1cm}
    \begin{minipage}{0.4\textwidth}
	\begin{lstlisting} [label={lst:repoSummaryy},language=C,caption={bce9420},mathescape=true]
- $Gemfile.lock$
+ $Gemfile.rails-4.0.x.lock$
+ $Gemfile.rails-4.1.x.lock$
+ active_scaffold (3.4.34)
        \end{lstlisting}
    \end{minipage}
    \hspace{1cm}
    \begin{minipage}{0.4\textwidth}
	\begin{lstlisting} [label={lst:repoSummaryz},language=C,caption={87496cf},mathescape=true]
$Gemfile.rails-4.0.x.lock,$
$Gemfile.rails-4.1.x.lock$
- active_scaffold (3.4.34)
+ active_scaffold (3.4.35)
        \end{lstlisting}
    \end{minipage}
    \caption{The git diffs for a selection of the commits from Fig.~\ref{fig:activescaffold-hist}. }
    \label{fig:activescaffold-code}
\end{figure}
}

In the commit 10d78ad in Fig.~\ref{fig:activescaffold-hist}, the author upgraded the version number of the package from 3.4.34 to 3.4.35 in the \texttt{version.rb} file~\cite{activescaffold-break}. 
However, in order to correctly bump the version number in this project, a user needs to change the number in both the \texttt{version.rb} file, and a \texttt{Gemfile.lock} file.
The same user fixes this error in commit a086ca7 so that the repository is again passing the build (cf. annotation  
$\textit{fix}_1$  on the edge). 

At the same time, another user submitted a pull request with the old version which split the \texttt{Gemfile.lock} into two separate files for different versions of the library rails.
In the merge process, the user inadvertently removed the original \texttt{Gemfile.lock} without coping over the fix for the version number.
After the merge process, the repository has ended up in commit bce9420 and is in a similar state to the previous broken commit 10d78ad.
There are now two files that are not consistent with \texttt{version.rb}.

In addition to global rules, \app is also learning local rules specific to each repository. From the prior experience with 
$\textit{fix}_1$, \app has learned
that the \texttt{PATCH} value in the \texttt{version.rb} is potentially problematic in this situation. Running \app at this point
 (commit bce9420), generates the error below:

\begin{lstlisting}[language=C, xleftmargin=.01\textwidth]
Predicted build failure based on potential error locations:
lib/active_scaffold/version.rb:Line 5
   PATCH
Multiple Files:Multiple Lines
  rails
active_scaffold.gemspec: Multiple Lines
  s.add_runtime_dependency
\end{lstlisting}

The error states that \app has predicted a build failure, and provides the explanation that this failure is likely due to one of the three listed keywords.
The next day the user discovered on their own that this is in fact the issue, and applied $\textit{fix}_2$ to bring the repository back to a correct state~\cite{activescaffold-fix}.

These two examples are just an illustration of the kind of errors \app can detect and predict.
In our evaluation, we tested our tool on more than 30 different larger scale GitHub repositories. Our average prediction rate is $83\%$ (cf. Sec.~\ref{sec:eval}).
 
\section{Preliminaries}

In this section we introduce the basic vocabulary in the continuous integration paradigm. We describe a formalism that we use to model the CI terms. This formalism is the basis for the future analysis and the rule derivation process.

\subsection{Repository Status} 

The main structures used in continuous integration testing are repositories. A repository contains all information required for a CI build, such as source code, automated tests, lists of library imports, build scripts, and similar files. We simply use a letter $\repoCode{}$ to denote a repository. We also use the superscript $\repoCode{}^{\textit{name}}$ as an optional name tag to reference a specific repository.

The continuous integration process monitors how a repository evolves
over time, so we use $\repoCode{t}$ to denote the state of repository $\repoCode{}$ at time $t$.
While many version control systems allow for a branching timeline (called a \textit{branch} in git), we 
linearize and discretize that timeline according to the build order in the CI tool. Let $\mathcal{L}$ be a linearization function.  
For simplicity of a notation, we use $\repoCode{t}$ instead of $\repoCode{\mathcal{L}(t)}$. This way the time indexes are
always non-negative integers.

Given a repository \repoCode{t}  a typical CI tool results in a build status \repoStatus{t}.
The exact definition of a ``status'', depends on the specific continuous integration tool. 
We generally categorize the possible status results as one of three options: \begin{itemize}
\item the repository is passing all tests, denoted by \repoStatus{t} = $Pass$ 
\item some tests in the repository have failed,  denoted by \repoStatus{t} = $Fail$
\item an error in the build process before tests can even be executed, denoted by \repoStatus{t} = $Err$
\end{itemize}

In this work we are specifically interested in the configuration of the repository and formally verifying its correctness, and not in the test cases themselves. For this reason, 
we only distinguish between the statuses where the build process was successful (relabelling both $Pass$ and $Fail$ as $Pass$) and states where the build process did not succeed (the $Err$ status).

Of a particular interest are changes in the repositories that cause the status to change, for example when the repository status changes from passing to erroring. To capture that, we slightly abuse the  \repoStatus{} function and for brevity, we introduce the following notation:\begin{align*}
   \repoStatus{t,t+1} = PE :\Leftrightarrow \repoStatus{t}=Pass \land \repoStatus{t+1}=Err\\
   \repoStatus{t,t+1} = EP :\Leftrightarrow \repoStatus{t}=Err \land \repoStatus{t+1}=Pass
\end{align*}

\subsection{Repository Summary}
\label{sec:prelim-summary}

We now describe how we model repositories so that we can easily reason about them. Every repository has a very 
non-homogeneous structure, containing a number of files that are not relevant for deriving the properties about the 
CI configuration. Examples of these files include ``readme'' files, \texttt{.csv} files, or images. We first filter out all such files.
The filtering is based on the file extensions and is hard-coded in \app. Clearly, it can be easily changed to 
add or remove different file types. For example, for Ruby programs we consider all {\textit{*.rb}},
{\textit{Gemfile}}, {\textit{gemspec}} files. Our goal is to translate a repository to a single representation of 
the various syntax found in real world repositories. This will be an intermediate representation of the repository 
on which the learning process is built.

Given a repository $\repoCode{}$, let $\mathcal{F}(\repoCode{})$ be a ``filtered' repository containing only the 
relevant files. Each $\mathcal{F}(\repoCode{})$ can be see as a union of strings (lines in the files). There are a couple of parsing 
issues, such as when a command spans multiple lines, but these issues are technical and not relevant to this 
presentation. It suffices to say that in our implementation we address these technical details so that the resulting repository is indeed a union 
of relevant text lines:  $\mathcal{F}(\repoCode{}) = \cup_{j=0}^m l_j$.

We introduce the abstraction function $\alpha$ which takes a repository $\mathcal{F}(\repoCode{})$  and returns 
an abstracted version of $\mathcal{F}(\repoCode{})$ which is then amenable for further processing. The abstraction 
function $\alpha$ maps the raw source of all components in the repository that may influence the build 
status to a more uniform, intermediate representation. The abstraction $\alpha$
takes two arguments as input: a repository and a set of so-called \textit{code features extractors}, which we denote with $CF$. 
Broadly speaking, it is a set of functions, which define how to translate a line in repository 
into a tuple of consisting of a keyword and a numerical value. 
Since we use a machine learning approach (\eg, decision trees) for building a verification model, we must map all of relevant keywords to a vector of reals, $\reals^{n}$.
Additionally, since we will later need to extract a human readable explanation for out models classifications, the feature vector we create must also have annotations so that each $CF$ produces a $(String,\reals)$.
Given the set of code features extractors $CF = \cup_{i=1}^n f_i$, the abstraction function is defined as follows:
\begin{align*}
  & \alpha (\mathcal{F}(\repoCode{}), CF) = \alpha (\cup_{j=0}^m l_j, CF) = \cup_{j=0}^m \tilde{\alpha} (l_j, CF)\\
  & \tilde{\alpha} (l, \cup_{i=1}^n f_i) = \cup_{i=1}^n f_i(l)
\end{align*}

Given a repository $\repoCode{t}$, with $\repoSummary{t}$ we denote the result of applying the abstraction $\alpha$ 
to $\repoCode{t}$. We call $\repoSummary{t}$  the {\emph{repository summary}} and it is a vector of  $(String,\reals)^{|CF|}$.

{
\begin{figure}[!h]
    \centering
    \begin{minipage}{.42\textwidth}
	\begin{lstlisting} [label={lst:repoCode1},language=C,caption={\repoCode{4}},mathescape=true]
 	 import Tweet V1.0
 	 import RndMsg V2.0

	 msg = RndMsg()
	 tweet(msg)
	\end{lstlisting}
    \end{minipage}
    \hspace{1cm}
    \begin{minipage}{0.42\textwidth}
	\begin{lstlisting} [label={lst:repCode2},language=C,caption={\repoCode{6}},mathescape=true]
	 import Tweet V2.0
	 import RndMsg V2.0

	 msg = RndMsg()
	 sendTweet(msg)
        \end{lstlisting}
    \end{minipage}
    %\hspace{1cm}
    \begin{minipage}{0.42\textwidth}
	\begin{lstlisting} [label={lst:repoSummary1},language=C,caption={\repoSummary{4}}]
	  (Tweet, 1.0)
	  (RndMsg, 2.0)
	  (tweet/sendTweet, -1)
	\end{lstlisting}
    \end{minipage}
    \begin{minipage}{0.42\textwidth}
	\begin{lstlisting} [label={lst:repoSummary2},language=C,caption={\repoSummary{6}}]
	  (Tweet, 2.0)
	  (RndMsg, 2.0)
	  (tweet/sendTweet,1)
	\end{lstlisting}
    \end{minipage}
    \caption{Two examples for extracting code features
(Listings~\ref{lst:repoSummary1} and~\ref{lst:repoSummary2}) from
repository code (Listings~\ref{lst:repoCode1} and~\ref{lst:repCode2} respectively) at time points 4 and 6 of a commit history}
    \label{fig:repoCode}
\end{figure}
}

Fig.~\ref{fig:repoCode} provides an example of the code feature extraction. In this example, in the learning 
process we use two main types of code feature extractors: \textit{magic constant code features} and \textit{diff 
code features}.

The magic constant code features tracks the use of hard-coded numerical constants in a repository.
These code features are especially important in CI configurations as they usually specify library version constraints.
In Fig.~\ref{fig:repoCode}, we extract the magic constant code features of the versions of the libraries.
Although the implementation may vary, the core idea that if a line is prefixed with a known keyword, we build a 
tuple, where the first element is that known keyword, and the second element in the tuple is the remaining constant 
value.

Diff code features are built from the diffs between two commits. The goal is to capture what was newly added or 
removed from repository. 
For two keywords $k_1$ and $k_2$ (\lstinline{tweet} and \lstinline{sendTweet} in 
Fig.~\ref{fig:repoCode}), the line abstraction 
function $\tilde{\alpha}$ will compare those keywords relative to two repositories \repoCode{i} and \repoCode{j}. 
If keyword $k_1$ is used in \repoCode{i} but not \repoCode{j}, $\tilde{\alpha}$ returns $(k_1 / k_2,-1)$. 
If keyword $k_2$ is used in \repoCode{j} but not \repoCode{i}, $\tilde{\alpha}$ returns $(k_1 / k_2, 1)$. 
If a repository \repoCode{k} does contain neither $k_1$ nor $k_2$, $\tilde{\alpha}$ returns $(k_1 / k_2, 0)$.
It cannot be that two repositories will contain both keywords, since otherwise they would not be the result of 
 the diffs between two commits.
More about learning diff code features and how they evolve over the time is given in Sec.~\ref{sec:gvl_learn}. 

As with many feature extraction tasks, code feature extraction must use
a set of templates built by an expert, a process known as ``feature
engineering''~\cite{scott99feature}.
We have defined some basic templates here, but combine these general templates with an automated feature extraction process in the learning process (cf.  Sec.~\ref{sec:gvl_learn}).
Automated code feature extraction was studies in~\cite{chae2017automatically}, but it was in the context of C-like languages and we could not reuse their results, because their language has much more structure than CI configurations.

\section{Overview}

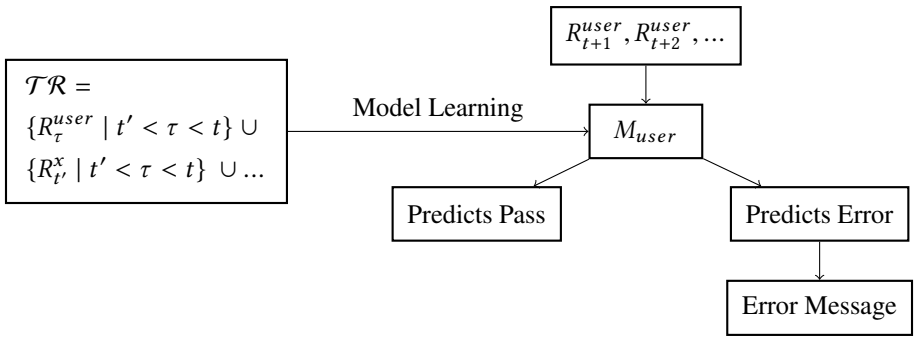
\begin{figure}[h!]
  \centering
  \begin{tikzpicture}
\tikzstyle{b} = [rectangle, draw, node distance=0.5cm, minimum width=1.5cm, minimum height=2em, thick, align=center, inner sep=0.2cm
                , execute at begin node={\begin{varwidth}{15em}}
                , execute at end node={\end{varwidth}}]
\tikzstyle{c} = [fill=gray!30, rounded corners=5pt]
\tikzstyle{l} = [draw, -latex',thick]

    \node [b] (repo) {$\begin{aligned}
				& \trainingSet  = \\ 
				& \{\repoCode{\tau}^{user} \ |\ t' < \tau < t \}\ \cup \\ 
				& \{\repoCode{t'}^{x}      \ |\ t' < \tau < t \}\ \cup ...
			\end{aligned}$ };
    \node [b, right=4cm of repo] (model) {$M_{user}$};
    \node [b, above=of model] (newcommit) {$\repoCode{t+1}^{user}, \repoCode{t+2}^{user}, ...$};
    \node [b, below left=of model] (resultP) {Predicts Pass};
    \node [b, below right=of model] (resultE) {Predicts Error};
    \node [b, below =of resultE] (errorMsg) {Error Message};

    %\draw[] 
    \path[->] 
	(repo) edge ["Model Learning"] (model)
    	(newcommit) edge (model)
        (model) edge (resultP)
        (model) edge (resultE)
        (resultE) edge (errorMsg);
\end{tikzpicture}
  \caption{\app works in two stages: in the training stage we learn a model specific to a user's repository, then in the verification stage we use that model to provide feedback on new commits.}
  \label{fig:high-overview}
\end{figure}

To statically verify that a CI configuration file is correct, we would need to have an equivalent to the specification. 
Our project can also be seen as the process of automatically deriving a specification for CI configuration files. We analyze 
a large number of existing CI configuration files and corresponding repositories and we  
build a model, represented as a set of rules, describing the properties that have to hold on those files.
This is also different from a traditional verification process because our specification will be learned from existing files, 
and thus (potentially) incomplete. Once we have a specification, 
the verification process then predicts the build status of the repository, and provide an explanation to the user for the 
prediction.
A high level overview of the flow of our method is shown in Fig.~\ref{fig:high-overview}.

A user first provides a training set, denoted by \trainingSet, of commits from a number of repositories. This training set 
should include the specific repository that the user is working on, $\repoCode{}^{user} \in \trainingSet$. The training set
can also contain different repositories, $\repoCode{}^{x}$.
Our tool will use this training set to produce a model specific to the user's repository, $M_{user}$.
This model can be used to check the correctness of future CI configurations when a user pushes a commit in $\repoCode{}^{user}$.
If the model predicts a failing build, it also produces an error message to help the user identify the source of the potential failure.

Existing efforts on prediction of build failures for CI can achieve relatively high accuracy (75\%-90\%)
but they rely on metadata in the learning process~\cite{ni2017cost,wolf2009predicting,hassan2006using}. These metadata 
include for example, a numbers of user commits, or a number of lines of code changed, or which files are changed. None of these approaches looks directly into user programs and configuration scripts.
While the metadata appears to be useful for prediction, it cannot not provide the user with any information that can help them isolate the root cause of the error.
In our work, in addition to build a failure prediction system based on the data analysis, we also provide users with  the suspected cause of failure, so that the user might proactively fix that error.

The goal of producing useful error messages creates two addition challenges:
\begin{itemize}
\item {\textbf{Finding the right set of code code feature}}  Learning a classification model depends on having
an efficient code feature extraction method that only uses a set of code features that are directly tied to the code that a 
user writes. This restricts us from using any kind of metadata in our learning process, since metadata cannot be changed by a 
user in a single commit.
\item {\textbf{Explainable learning}} We need to extend our learning strategy so that it also produces justification for the 
classification. This problem is commonly referred to as the issue of \textit{legibility} in
machine learning~\cite{forman2003extensive}. In the recent years it gain again popularity under the name \textit{explainable 
AI}~\cite{gunning2017explainable}.
\end{itemize}

\iffalse

In addition to learning a classification model, we need to first find a code feature extraction method ($\alpha: (\repoCode{},\{CF\}) \to \repoSummary{}$) that only uses a set of code features $\{CF\}$ that are directly tied to the code a user can write.
This restricts us from using any kind of metadata in our learning process, since metadata cannot be easily changed by a user in a single commit.
Second, we need to use a learning strategy that produces justification for the classification.
This is commonly referred to as the issue of \textit{legibility} in
machine learning~\cite{forman2003extensive}, and also as \textit{explainable AI}~\cite{gunning2017explainable}.

\fi

\section{Learning}
\label{sec:learning}

\begin{figure}[h!]
  \centering
  \begin{tikzpicture}
\tikzstyle{b} = [rectangle, draw, node distance=0.5cm, minimum width=1.5cm, minimum height=2em, thick, align=center, inner sep=0.2cm
                , execute at begin node={\begin{varwidth}{15em}}
                , execute at end node={\end{varwidth}}]
\tikzstyle{c} = [fill=gray!30, rounded corners=5pt]
\tikzstyle{l} = [draw, -latex',thick]

    \node [b,c] (repo) {Training Set};
    \node [b, right=of repo] (features) {Code Feature \\ Extraction};
    \node [b, right=of features] (abr) {Abstraction \\ Based \\ Relabelling};
    \node [b, right=of abr] (m) {Learning Module};
    \node [b,c, right=of m] (model) {Learned \\ Classifier};
    \node [b, right=of repo] (features) {Code Feature \\ Extraction};
    \node [b, node distance=1cm, below=of m] (misc) {Misclassification \\ Ranking};
    \node [b, node distance=0.9cm, below=of features] (gen) {Code Feature \\ Extraction Refinement};

    \draw [l] (repo) -- (features);
    \draw [l] (features) -- (abr);
    \draw [l] (abr) -- (m);
    \draw [l] (m) -- (misc);
    \draw [l] (m) -- (model);
    \draw [l] (misc) -- (gen);
    \draw [l] (gen) -- (features);
\end{tikzpicture}
  \caption{The MiGAR loop in context of learning Continuous Integration specifications.}
  \label{fig:build}
\end{figure}
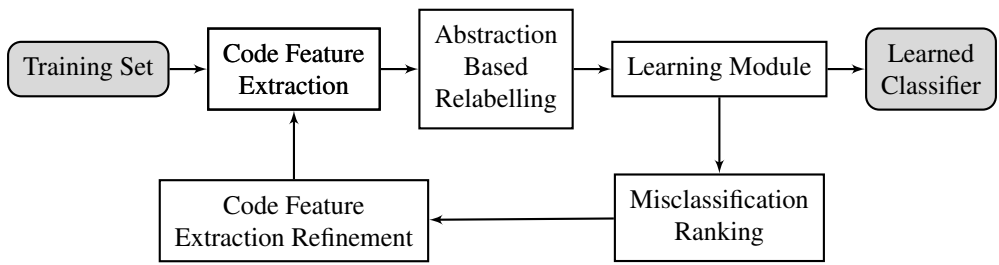

In order to learn both the set of code feature extractors and a model of configuration correctness, our system uses a refinement loop as shown in Fig~\ref{fig:build}, which is inspired by counterexample guided abstraction refinement (CEGAR)~\cite{clarke2000counterexample}.
The CEGAR loop has been used for various model checking tasks~\cite{BeyerModelChecking, BallSLAM, Tian2017}. A similar loop is also widely 
used in program synthesis~\cite{SolarCEGIS,JhaCEGIS14a}.
As we have adapted the CEGAR loop to the context of machine learning, we do not use counterexamples, but rather misclassifications as the driver behind our abstraction refinement.
This is critical difference, as the information obtained from a counterexample is much stronger than the information gained from a misclassification.
We will reference our algorithm as a MiGAR (Misclassification guided abstraction refinement) in the following sections.

The MiGAR loop helps us automate feature extraction from repository code with feedback from our learning mechanism.
Similar to the CEGAR loop which usually starts with a weak abstraction (such as loop invariants) and gradually strengthens the invariant during the loop~\cite{nguyen2017counterexample,greitschus2017loop}, our MiGAR loop starts with an initial feature abstraction that captures only a few features, and gradually refines the abstraction to include more features with each loop.

The main issue that the MiGAR loop addresses is the need to extract enough code features to have accurate classification, but not so many that we encounter issues with overfitting.
If the code feature extraction includes too few features, the learning step will be unable to find a model that can correctly classify a commit with its potential status.
On the other hand, if code feature extraction includes too many features, we will run into an overfitting issue, where the learning module will not be able to generalize to new commits.
We would like to select the highest level of abstraction (code features) possible that can capture the differences between passing and failing - the less abstract the code features used, the less likely the results of learning are likely to generalize.
The MiGAR loop act as an incremental automated feature selection method that balances these two problems.
As we will explain in the following sections, this approach is specialized for the situation where out training set of code has incremental changes as result of the commit history of the version control.

\subsection{Global vs Local Learning}
\label{sec:gvl_learn}

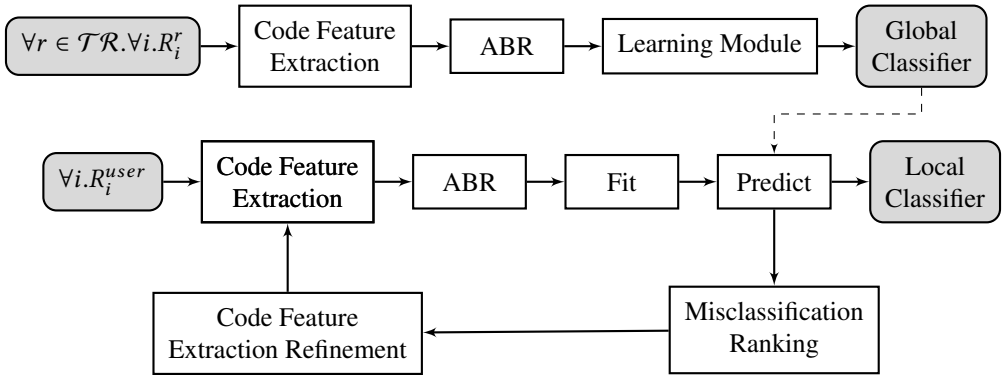
\begin{figure}[h!]
  \centering
  \begin{tikzpicture}
\tikzstyle{b} = [rectangle, draw, node distance=0.5cm, minimum width=1.5cm, minimum height=2em, thick, align=center, inner sep=0.2cm
                , execute at begin node={\begin{varwidth}{15em}}
                , execute at end node={\end{varwidth}}]
\tikzstyle{c} = [fill=gray!30, rounded corners=5pt]
\tikzstyle{l} = [draw, -latex',thick]
\tikzstyle{ll} = [draw, -latex',dashed,]

    % Global learning
    \node [b,c] (repoG) {$\forall r \in \trainingSet. \forall i. \repoCode{i}^{r}$};
    \node [b, right=of repoG] (featuresG) {Code Feature \\ Extraction};
    \node [b, right=of featuresG] (abrG) {ABR};
    \node [b, right=of abrG] (mG) {Learning Module};
    \node [b,c, right=of mG] (modelG) {Global \\ Classifier};

    \draw [l] (repoG) -- (featuresG);
    \draw [l] (featuresG) -- (abrG);
    \draw [l] (abrG) -- (mG);
    \draw [l] (mG) -- (modelG);

    % Local Learning
    \node [b,c, node distance=1cm,below=of repoG] (repo) {$\forall i. \repoCode{i}^{user}$};
    \node [b, right=of repo] (features) {Code Feature \\ Extraction};
    \node [b, right=of features] (abr) {ABR};
    \node [b, right=of abr] (fit) {Fit};
    \node [b, right=of fit] (pred) {Predict};
    \node [b,c, right=of pred] (model) {Local \\ Classifier};

    \node [b, right=of repo] (features) {Code Feature \\ Extraction};
    \node [b, node distance=1cm, below=of pred] (misc) {Misclassification \\ Ranking};
    \node [b, node distance=0.9cm, below=of features] (gen) {Code Feature \\ Extraction Refinement};
    
    \draw [l] (repo) -- (features);
    \draw [l] (features) -- (abr);
    \draw [l] (abr) -- (fit);
    \draw [l] (fit) -- (pred);
    \draw [->,ll] (modelG) to[|-|] (pred);
    \draw [l] (pred) -- (misc);
    \draw [l] (pred) -- (model);
    \draw [l] (misc) -- (gen);
    \draw [l] (gen) -- (features);
\end{tikzpicture}
  \caption{To learn rules for the repository $\repoCode{}^{user}$ we combine a MiGAR loop for local rules with an additional step based on a static, initial code feature extraction for global rules.}
  \label{fig:overview}
\end{figure}

In our manual study of the errors that contribute to failed CI builds, we noticed that there were some errors that were common across multiple repositories, and others that were highly dependent on the specific repository being examined.
For example, library version incompatibilities are a common constraint
that globally holds across all repositories (\eg, no matter what, libA V1 cannot be used with libB V2).
On the other hand, some repository might define its own API and implicitly impose a constraint local to that repository that the function \texttt{f} cannot be called in the same file as the function \texttt{g}.
In order to capture this observation, we separate our learning process into two stages, shown in Fig.~\ref{fig:overview}.

In the first stage, we learn \textit{global rules}, which are derived from the full set of available repositories in the training set $\trainingSet$.
When learning over many repositories to generate global rules, we use an initial code feature extraction based on \textit{magic constant code features}, as described in Sec~\ref{sec:prelim-summary}.
When trying to extract all magic constant code features from all repositories in a training set, we quickly obtain a very large number (depending on the exact coding style used in the repository).
This presents a challenge in scalability of learning, but also can result in overfitting, as the learning algorithm can rely on unique magic constants to identify repository state.
To prevent this situation, we require each code feature $cf$ to pass a certain threshold of \textit{support} from the training set to be considered in the set of code feature extractors $CF$ use in the abstraction function $\alpha$.
Support is a metric used in other knowledge discovery algorithms, such as association rule learning~\cite{agrawal1993mining}.
We say that a code feature $cf$ has $support(cf)$ as defined by how many commits in the training set contains a particular code feature, with respect to the size of the training set \trainingSet, where we define $|\trainingSet|$ to be the total number of commits over all repositories.
The appropriate level of support must be determined by the user according the content of the training set, though the typical starting value used in association rule learning is usually 10\%.

\begin{equation*}
 \textit{support(cf)} = \frac{|\{\forall x \in \trainingSet.\ \forall i.\ \repoCode{i}^{x} \mid cf \in \repoSummary{i}^{x}\}|} {|\trainingSet|}
\end{equation*}

In the second stage, we learn \textit{local rules}, which are derived from a specific repository.
When learning over a single repository to generate local rules (the bottom half of Fig.~\ref{fig:overview}), we use the same initial code feature extraction as for the global rules based on magic constants.
However, for local rules, we learn the initial abstraction with respect to only $\repoCode{}^{user}$ for finding keywords and calculating support.
We then use a MiGAR loop to refine this abstraction, and with each iteration of the loop we refine the set of features by adding new \textit{diff code features}, as described in Sec~\ref{sec:prelim-summary}.

In learning local rules, we use the learned global classifier as an oracle during the prediction process. 
During the MiGAR loop, we refine the feature extraction on repository states that were misclassified by the local model.
However, as we have already created a global model, we want to avoid wasting cycles of the MiGAR loop to relearn global rules on a local level.
Additionally, the MiGAR loop only refines extraction on diff code features, 
  and if we had already learned that the root cause of an error was in a magic constant code feature from the global classifier, 
  forcing the local model to try to learn classification based on diff code features is unlikely to produce a useful result.

In order to generate a new local code feature, we need to select a pair of \repoCode{t} and \repoCode{t+1} from which we will build our diff code feature.
To select the commit that is most likely to help us improve our learned model and correctly classify a repository state, we use a ranking algorithm.
Our ranking algorithm first filters commits for patterns where $\repoStatus{t,t+1}=PE$ or $EP$, then ranks these commits by increasing size of each commit.
The intuition here is that the most important features are likely to be found when a small change resulted in a status change of the repository.
Furthermore, we can rank commits by size, with the intuition that when a small commit changes a build status, that commit contains more impactful and information-rich changes. 
This ranking algorithm can be tuned to the specific application, and include more domain specific information as it is available.

Once we have built both a global and local classifier we can combine the prediction results of these two classifiers to make predictions with more confidence.
We have three possible cases in combining global and local classifiers: either the two models are in agreement, or the global model predicts $E$ and the local model predicts $P$, or the local model predicts $E$ and the global model predicts $P$.
In the case that the two models are in agreement, we take the agreed upon classification.
When resolving conflicting classifications, the method can depend on the overall design goal of the tool.
In deploying verification tools based on machine learning, it is often most important to reduce false positives (reporting an error when there is no problem), in which case a conservative measure is to reject a error report when either model classifies the repository as safe.
Further refinements can be made based on design goals and domain specific assumptions about the accuracies of the global and local models.

\subsection{Abstraction Based Relabelling}

One of the key challenges in synthesizing specifications for real world systems is the deep level of dependencies and many points of possible failure.
CI builds are not pure in that sense that some CI builds can fail due to a dependency of the repository failing, a network outage, or even a CI hardware failure.
This non-determinism of configuration behavior has also been observed in other configuration domains~\cite{ShambaughWG16}.
This presents a challenge for learning since it is not always possible to trace the source of that error to code features in the \repoSummary{}, which results in a noisy training set (two identical data points map to different classifications).
In order to reduce the noise that our learning module needs to handle, we want to ensure  $\repoSummary{i} = \repoSummary{j} \implies \repoStatus{i} = \repoStatus{j}$.

In order to achieve this property of our training set, we introduce a step into our framework we call \textit{Abstraction Based Relabelling} (ABR).
In ABR, we examine the timeline of the representation of each repository state as code features. 
If at some point the status changes, but the code features do not, we relabel that erroring status as a passing state.
\begin{align*}
\text{If } \repoStatus{t,t+1} = PE \land \repoSummary{t} = \repoSummary{t+1}
\text{, then relabel } \repoSummary{t+1} = P \\
\text{If } \repoStatus{t,t+1} = EP \land \repoSummary{t} = \repoSummary{t+1}
\text{, then relabel } \repoSummary{t} = E
\end{align*}

\newcolumntype{C}{>{$}c<{$}} % math-mode version of "l" column type

\bgroup
{\setlength{\tabcolsep}{0.37cm}
\renewcommand{\arraystretch}{1.2}
\begin{table}[htbp]
\centering
\caption{An example learning process demonstrating ABR and a refinement step for new code feature generation. Source code for \repoSummary{4} and \repoSummary{6} are listed in Fig.~\ref{fig:repoCode}.}
\label{table:ABR}
\begin{tabular}{|l||C|C|C|C|C|C|}
\hline
Code Features
  & $\repoSummary{1} $
  & $\repoSummary{2} $
  & $\repoSummary{3} $
  & $\repoSummary{4} $
  & $\repoSummary{5} $
  & $\repoSummary{6} $\\
\hline \hline
Original Status & P & E & P & E & E & P \\ 
\hline \hline
Initial Abstraction    &&&&&&  \\ \hdashline
\quad $\repoSummary{}$ &&&&&&  \\
\quad \quad import Tweet   & 1.0 & 1.0 & 1.0 & 1.0 & 2.0 & 2.0   \\
\quad \quad import RndMsg  & 1.0 & 1.0 & 1.0 & 2.0 & 2.0 & 2.0  \\
ABR Status      & P & \bf{P} & P & E  & \bf{P} & P\\ 
\quad $\repoModel{}$ &&&&&& \\
\quad \quad Tweet == RndMsg & P & P & P & E & P & P \\
\hline  \hline
Refined on $\repoCode{6}$ &&&&&&  \\ \hdashline
\quad $\repoSummary{}$ &&&&&&   \\
\quad \quad tweet/sendTweet & 0 & 0 & 0 & -1 & -1 & 1 \\ 
ABR Status      & P & \bf{P} & P & E & E & P  \\ 
\quad $\repoModel{}$ &&&&&&  \\
\quad \quad Tweet == RndMsg $\land$ &&&&&& \\
\quad \quad FB $\geq$ RndMsg & P & P & P & E & E & P  \\
\hline 
\end{tabular}
\end{table}
\egroup

To understand the motivation for, and effect of ABR, we provide an example learning process in Table~\ref{table:ABR}.
In this example, $\repoStatus{2}=E$ because of a network failure, while $\repoStatus{5}=P$ was due to the fact that the Tweet library depreciated the function \texttt{tweet} and replaced it with \texttt{sendTweet}.
During our learning process we do not have information on the root cause of these failures, but we can observe the code features and CI status.
In the initial abstraction, we only have the magic constant code features and relabel both \repoStatus{2} and \repoStatus{5} accordingly.
After building a model, we may want to increase our accuracy and so enter into a MiGAR loop where we extract the diff code feature \texttt{tweet}/\texttt{sendTweet}.
We now have distinguishing features between \repoSummary{5} and \repoSummary{6}, so we no longer relabel \repoStatus{5} in the ABR step and the learning module can build a model that captures these features.

\subsection{Decision Trees}

Selecting an appropriate learning module is key to ensuring that our system can provide justification for classification results.
In this work we have used decision tree learning because of the transparent structure of the learned model.
A decision tree consists of decision nodes, which contain rules to dictate classification, and end nodes which denote the final decision on the classification.
A justification for a decision tree's classification is easily obtained by tracing the \textit{decision path} through the tree for a given sample.
An example of a decision tree that might be learned by taking the repositories in Table~\ref{table:ABR} as a training set is given in Fig~\ref{fig:dtree}.
In Fig.~\ref{fig:real_dtree}, to give a sense of scale, we provide a rendering of an actual decision tree our tool, \app, generated.

\begin{figure}[h!]
 \centering
    \begin{minipage}{.4\textwidth}
      \begin{tikzpicture}
\tikzstyle{b} = [rectangle, draw, node distance=0.5cm, minimum width=1.5cm, minimum height=2em, thick, align=center, inner sep=0.2cm
                , execute at begin node={\begin{varwidth}{15em}}
                , execute at end node={\end{varwidth}}]
\tikzstyle{c} = [fill=gray!30, rounded corners=5pt]
\tikzstyle{l} = [draw, -latex',thick]

    \node [b] (root) {Tweet=1};
    \node [b, below left=of root] (rnd1) {RndMsg=1};
    \node [b, below right=of root] (tweet2) {Tweet=2};
    \node [b, below left=of tweet2] (rnd2) {RndMsg=2};
    \node [b,c, below left=of rnd2] (d1) {Error};
    \node [b,c, below right=of rnd2] (d2) {Pass};

    \coordinate [below left=0.2cm of rnd1] (t2l);
    \coordinate [below right=0.2cm of rnd1] (t2r);
    \path[->] 
	(root) edge ["True"] (rnd1)
    	(root) edge ["False"] (tweet2)
    	(root) edge (tweet2)
    	(tweet2) edge (rnd2)
    	(rnd1) edge (t2r)
    	(rnd1) edge (t2l)
    	(rnd2) edge (d1)
    	(rnd2) edge (d2);
\end{tikzpicture}
      \caption{An example of a decision tree with white boxes as decision nodes and gray boxes as end nodes.}
      \label{fig:dtree}
    \end{minipage}
    \hspace{1cm}
    \begin{minipage}{0.4\textwidth}
      \includegraphics[width=0.8\textwidth]{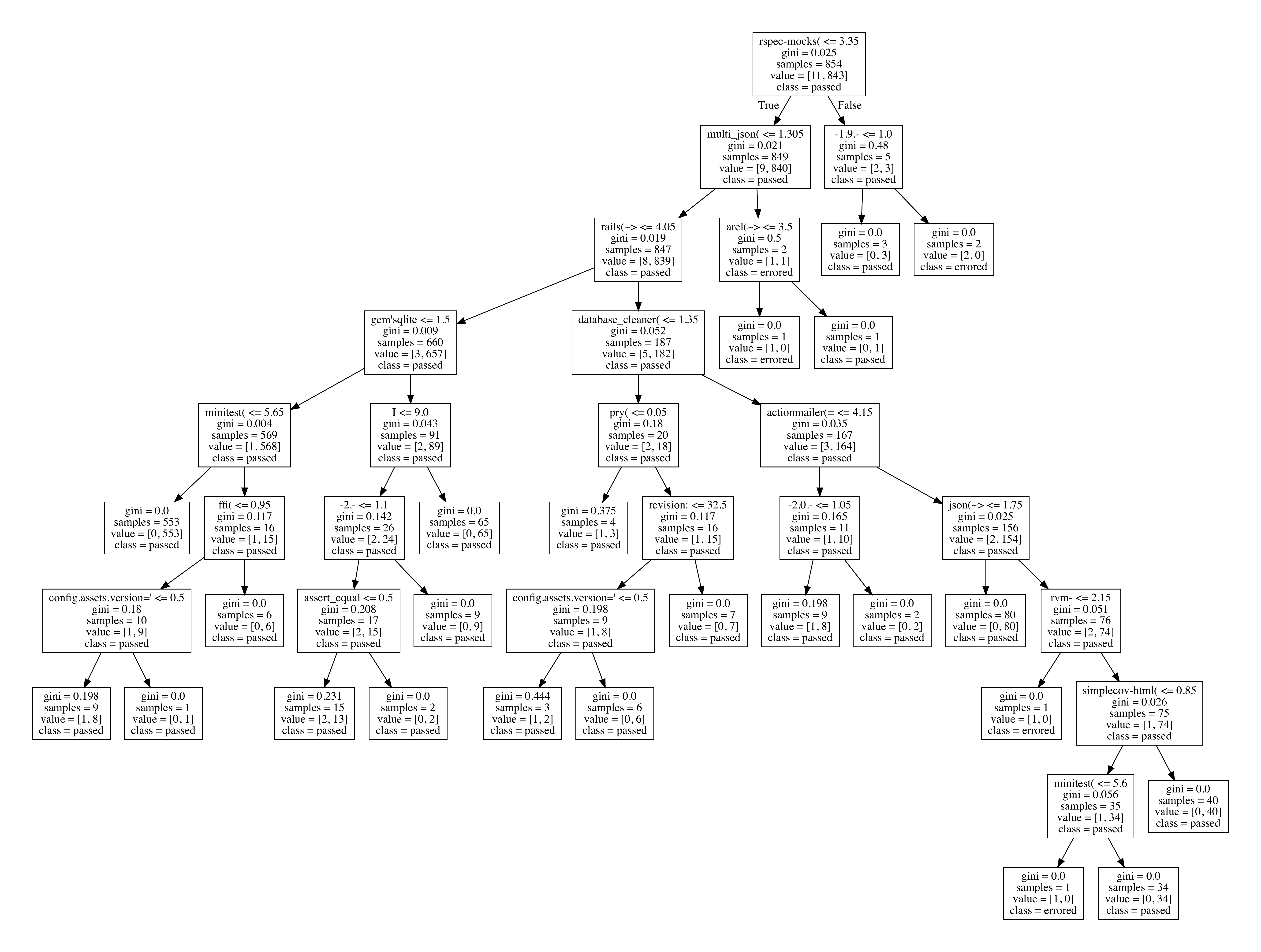}
      \caption{An actual tree generated by our tool, \app.}
      \label{fig:real_dtree}
    \end{minipage}
\end{figure}

One of the caveats of using decision trees is the need for a balanced training set~\cite{chawla2004special}.
This is particularly important for detecting rare events, such as in our case, where we detect potential misconfigurations that will lead to a CI build error.
If the training set contains 99\% passing status, and only 1\% erroring status, a learning algorithm can easily achieve 99\% accuracy by simply classifying all repository states as passing.

In order to ensure useful learning in the presence of rare events, we need to handle an imbalanced training set.
Random undersampling is a traditional approach for this, but it is also possible to use domain knowledge to make better undersampling choices.
In our domain of CI, we use a undersampling technique to remove sequences of $\repoStatus{}=P$ in a \trainingSet, while leaving all $PEP$ patterns.
We remove such sequences until the overall event rate of a $\repoStatus{}=E$ in \trainingSet is at least some threshold (in our case 30\% was sufficient).
This produces a training set that retains the most possible $\repoStatus{t,t+1,t+2}=PEP$ patterns, which yields diff code features that are more specific and more likely to be helpful.
Other techniques could be used such as random oversampling or synthetic minority oversampling (SMOTE)~\cite{chawla2002smote}, but these are again more generic and do not leverage the domain specific information that our samples have a historical order.

\section{Evaluation}
\label{sec:eval}

To evaluate the feasibility and effectiveness of our approach, we have implemented a tool, \app, which verifies repositories using the open-source continuous integration testing tool, TravisCI.
TravisCI is an ideal test bed, as the tool is free for open-source projects and widely used on GitHub.
The log history of the builds for a number of large, active, and open-source project has been made available through the TravisTorrent dataset~\cite{travis-torrent}.
This dataset directly provides us with the labeled, temporally ordered commit data over many repositories we need as a training set.

In our evaluation, we answer three key question that give an intuition of the behavior of \app.
First, we ask \textit{can \app accurately predict the build status before the build actually executes}? 
Second, since the goal of \app is not only to predict build status, but also to give a helpful explanation of potential fixes, we ask \textit{are the errors that \app reports correct}?
Lastly, our goal is to provide a tool that can be integrated into the workflow of a developer using CI in practice, so we ask \textit{is \app able to scale the learning process to large commit histories?}

\subsection{Accuracy of Prediction}

In order to evaluate the accuracy of \app in predicting build status, we have set our evaluation up to model the way \app would be used in the real world. 
In practice, \app would be trained over a set of commits up until the present moment, and used to predict commits for one day. 
Then, at the end of the day, \app would build a new model, incorporating the information from that day, so that the model the next day can learn from any mistakes that were made.
On a timeline, this looks as shown in Fig.~\ref{fig:timeline}

\begin{figure}[h!]
  \centering
  \begin{tikzpicture}
\tikzstyle{b} = [rectangle, draw, node distance=0.9cm, minimum width=1.5cm, minimum height=2em, thick, align=center, inner sep=0.2cm
                , execute at begin node={\begin{varwidth}{15em}}
                , execute at end node={\end{varwidth}}]
\tikzstyle{c} = [fill=gray!30, rounded corners=5pt]
\tikzstyle{l} = [draw, -latex',thick]

    \node [b] (10d7) {10d78ad \\ Jan 1};
    \node [b, right=of 10d7] (a086) {a086ca7 \\ Jan 2};
    \node [b, below=of a086] (9b09) {9b09ff3 \\ Jan 2};
    \node [b, right=of a086] (2941) {2914a53 \\ Jan 3};
    \node [b, below=of 2941] (84ea) {84eabc9 \\ Jan 3};
    \node [b, right=of 2941] (bce9) {bce9420 \\ Jan 3};
    \node [b, right=of bce9] (8749) {87496cf \\ Jan 4};

    \path[->] 
	(10d7) edge (a086)
    	(a086) edge (84ea)
        (a086) edge (2941)
        (9b09) edge (84ea)
        (84ea) edge (bce9)
        (2941) edge (bce9)
        (bce9) edge (8749);

    \coordinate[above right=0.5cm and 0.5cm of a086] (x1);
    \coordinate[below left=1.1cm of 9b09] (x2);
    \node [above right=0.1cm and -0.6cm of x2] (t1) {Training Set for Jan 3};
    \draw [->,dashed] (x1) |- (x2);

    \coordinate[above right=0.5cm and 0.5cm of bce9] (x3);
    \coordinate[below left=1.1cm and 0cm of 9b09] (x4);
    \node [above right=0.1cm and 3.6cm of x4] (t1) {Training Set for Jan 4};
    \draw [->,dashed] (x3) |- (x4);
\end{tikzpicture}
  \caption{\app rebuilds a model at the end of everyday to provide better prediction the next day}
  \label{fig:timeline}
\end{figure}
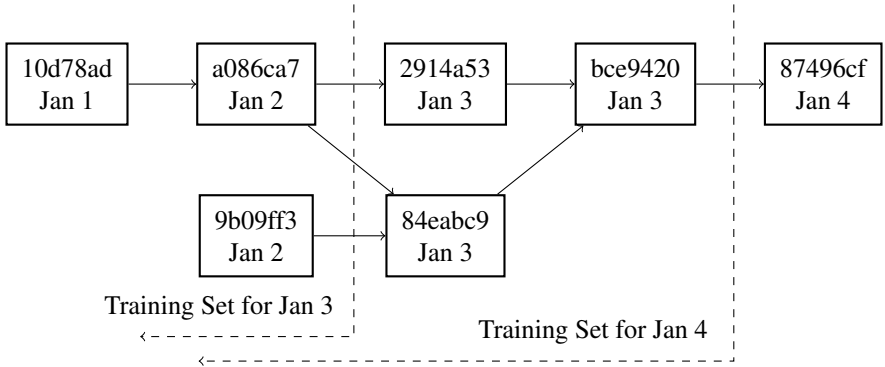

This is the strategy we employed in the evaluation presented in Table~\ref{table:accuracy}.
For a repository with $t$ total commits $(\repoCode{0},...,\repoCode{t})$, we build the first model with $t/2$ commits, and use that model to evaluate commit $\repoCode{(t/2)+1}$.
We then rebuilt the model with a training set of $(t/2)+1$ commits and used that model to predict commit number $\repoCode{(t/2)+2}$.
From this, we find that \app can accurately predict build status with an overall accuracy of $83\%$.
This is competitive with other tools for build status prediction based on metadata such as the committer's historical rate of correct commits, the size of the commit, sentiment analysis on the commit message, etc~\cite{ni2017cost,wolf2009predicting,hassan2006using,Paixao2017}.

In deploying probabilistic verification tools, one of the most important metrics is that the tool has a very low false positive rate.
In practice, this is the single metric that most often prevents adoption as users will quickly ignore tools with too many false alarms.
Table~\ref{table:accuracy} shows that the rate of false positive reports is 1:20.
Looking specifically at the false positive to true positive rate we report roughly one false positive for every two true positives.
While this certainly leaves room for improvement, it is a strong indication that we are on the right track.
The next step will be to run a user study to find the threshold of a false positives rate that users are willing to tolerate in CI verification.
A user study would be particularly interesting here as CI verification would occur much less frequently than, for example compiler errors, so users may have a relatively higher tolerance for false positives.
We leave such explorations to future work.

\bgroup
\def\arraystretch{1.3}
\begin{table}[t!]
\centering
\caption{Classification rates across 50 commits (pre-balancing of training set) of various repositories including number for TP (True Positive - correctly predicted error status), TN (True Negative), FP (False Positive - incorrect predicted error when actual status was pass), and FN (False Negative). We also report precision (= TP / TP + FP) and recall (= TP / TP + FN) with a '-' to indicate cases where this metric does not make sense (divide by zero).}
\label{table:accuracy}
\begin{footnotesize}
\begin{tabular}{|l|c|c|c|c|c|c|c|}
\hline
{\bf Repository Name} & {\bf Accuracy} & {\bf Precision} & 
{\bf Recall} & {\bf TP} & {\bf TN} & {\bf FP} & {\bf FN} \\ \hline \hline
\texttt{sferik/rails\_admin} & 88\% & 1.000 & 0.824 & 14 & 8 & 0 & 3  \\ \hline
\texttt{activescaffold/active\_scaffold} & 80\% & 0.750 & 0.429 & 3 & 17 & 1 & 4  \\ \hline
\texttt{fog/fog} & 88\% & 1.000 & 0.813 & 13 & 9 & 0 & 3  \\ \hline
\texttt{Homebrew/homebrew-science} & 92\% & 0.500 & 0.500 & 2 & 21 & 1 & 1  \\ \hline
\texttt{sferik/twitter} & 84\% & 0.000 & 0.000 & 0 & 21 & 1 & 3  \\ \hline
\texttt{ruboto/ruboto} &  72\% & 0.667 & 0.250 & 2 & 16 & 1 & 6  \\ \hline
\texttt{heroku/heroku} & 84\% & - & 0.000 & 0 & 21 & 0 & 4  \\ \hline
\texttt{honeybadger-io/honeybadger-ruby} & 76\% & - & 0.000 & 0 & 19 & 0 & 6  \\ \hline
\texttt{padrino/padrino-framework} & 84\% & 0.000 & 0.000 & 0 & 21 & 1 & 3  \\ \hline
\texttt{minimagick/minimagick} & 72\% & 0.333 & 0.167 & 1 & 17 & 2 & 5  \\ \hline
\texttt{colszowka/simplecov} & 100\% & - & - & 0 & 10 & 0 & 0  \\ \hline
\texttt{innoq/iqvoc} & 68\% & - & 0.000 & 0 & 17 & 0 & 8  \\ \hline
\texttt{openSUSE/open-build-service} & 84\% & 0.750 & 0.500 & 3 & 18 & 1 & 3  \\ \hline
\texttt{kmuto/review} & 92\% & 1.000 & 0.333 & 1 & 22 & 0 & 2  \\ \hline
\texttt{loomio/loomio} & 72\% & 0.250 & 0.200 & 1 & 17 & 3 & 4  \\ \hline
\texttt{rpush/rpush} & 88\% & 0.000 & 0.000 & 0 & 22 & 1 & 2  \\ \hline
\texttt{aws/aws-sdk-ruby} & 84\% & 0.000 & 0.000 & 0 & 21 & 1 & 3  \\ \hline
\texttt{test-kitchen/test-kitchen} & 76\% & 0.500 & 0.167 & 1 & 18 & 1 & 5  \\ \hline
\texttt{psu-stewardship/scholarsphere} & 80\% & 0.000 & 0.000 & 0 & 20 & 1 & 4  \\ \hline
\texttt{github/pages-gem} & 84\% & - & 0.000 &  0 & 21 & 0 & 4  \\ \hline
\texttt{supermarin/xcpretty} & 92\% & - & 0.000 & 0 & 23 & 0  & 2  \\ \hline
\texttt{SchemaPlus/schema\_plus} & 88\% & 0.500 & 0.333 & 2 & 20 & 2 & 1  \\ \hline
\texttt{poise/poise} & 96\% & - & 0.000 & 0 & 24 & 0 & 1  \\ \hline
\texttt{yuki24/did\_you\_mean} & 76\% & 0.500 & 0.167 & 1 & 18 & 1 & 5  \\ \hline
\texttt{fatfreecrm/fat\_free\_crm} & 90\% & 0.500 & 0.500 & 1 & 17 & 1 & 1  \\ \hline
\texttt{diaspora/diaspora} & 64\% & 0.250 & 0.143 & 1 & 15 & 3 & 6  \\ \hline
\texttt{garethr/garethr-docker} & 92\% & 1.000 & 0.500 & 2 & 21 & 0  & 2  \\ \hline
\texttt{cantino/huginn} & 72\% & 0.000 & 0.000 & 0  & 18 & 1 & 6  \\ \hline
\texttt{ledermann/unread} & 80\% & 0.000 & 0.000 & 0  & 20 & 1  & 4  \\ \hline
\texttt{Growstuff/growstuff} & 90\% & 0.500 & 0.500 & 1 & 17 & 1 & 1  \\ \hline
\texttt{Shopify/shopify\_api} & 84\% & 0.000 & 0.000 & 0  & 21 & 1 & 3  \\ \hline
\texttt{jbox-web/redmine\_git\_hosting} & 88\% & - & 0.000 & 0  & 22 & 0 & 3  \\ \hline
\texttt{plataformatec/devise} & 80\% & 1.000 & 0.167 & 1 & 19 & 0  & 5  \\ \hline
\hline
\texttt{Average} & 83.0\% & 0.458 & 0.203 & 1.515 & 18.515 & 0.788 & 3.424  \\ \hline

\end{tabular}
\end{footnotesize}
\end{table}
\egroup

One interesting note on the false negative rate is that we should expect this to always be some positive number.
Since our method only looks at the code inside the repository itself, and does not account for the ``impurity'' of builds resulting from errors in dependencies and network failure, we cannot expect that \app will ever be able to detect these types of errors.
To understand the frequency of these ``impure'' errors we report a small survey of the root cause of errors in repositories on GitHub using TravisCI in Table~\ref{table:PE_summary}.
Although our sample is too small for inferring a proper distribution, this does give a sense of the frequency in which \app ``should'' report false negative (we report the repository should pass, when in fact it errors).

For this survey of errors, we used the TravisTorrent~\cite{travis-torrent} tool to select $\repoStatus{t,t+1}=EP$ (errored to passing) status points.
We then categorized the reasons for each error into the categories listed in Table~\ref{table:PE_summary}.
These categories were based off the build log of the erroring repository, as well as the code that was changed to make the repository pass in the subsequent build.
The list of these errors, including a link to the specific commit within the repository and an explanation for the classification is available in the supplementary materials.

\bgroup
\def\arraystretch{1.2}
{\setlength{\tabcolsep}{2em}
\begin{table}[]
\centering
\caption{Frequency of error types for a random sample from the TravisTorrent database. The ``Error in dependency'' and ``Network Failure'' are impure errors and out of scope for our method to detect.}
\label{table:PE_summary}
\begin{tabular}{|l|c|}
\hline
{\bf Error Type}            & {\bf Number of occurrences}  \\ \hline \hline
Missing file          & 4     \\ \hline
Version Inconsistency & 40    \\ \hline
Syntax Error          & 10     \\ \hline
Source Code error     & 53    \\ \hline
\hline
Error in dependency   & 9     \\ \hline
Network Failure       & 14     \\ \hline
\end{tabular}
\end{table}
\egroup

\subsection{Accuracy of Error Messages}

We have shown that \app can predict build status with some accuracy, but the key innovation is that we can provide a justification, or explanation, for the classification that our model provides.
However, just providing any explanation is not satisfactory.
In particular, we would like the explanation provided to correspond to a potential change in the code base that could fix the erroring build status.
To evaluate whether \app provides useful error messages, we check if any of the keywords we presented in the error message appear in the difference between the erroring commit and the next passing commit.
If the user has changed a keyword that we had suggested in order to fix the repository's build status, it is evidence that we suggested a correct root cause of the error.

Using this metric, we found that over the repositories listed in Table~\ref{table:accuracy}, the error we reported corresponded to the change that user eventually made to fix the build our 48\% and 38\% of the time, for the global and local rules respectively.
The average number of keywords reported in an error message (the decision path depth) by \app was 4.51 for global and 2.42 for local.
This means that in the local case, 38\% of the time we correctly identified a critical keyword in a breaking commit with an average of only 2.42 guesses.
%TODO would be really nice to know the average size of the git diff that we were searching through as well to show how hard it is to find these exact keywords - maybe we can do this for camera ready (or more likely the next submission)

We note that this is a very rough metric that only gives a sense of the usefulness of our error messages.
It is possible that the although the user changed a keyword we flagged, that keyword was not in fact very important.
It is also the case that there may be multiple correct fixes for a problem, and while we suggested a correct one, the user decided to take another fix.
However, overall these are very encouraging results that show it is indeed possible to (and \app succeeds in doing so) pinpoint the root cause of a CI misconfiguration from a learning approach.

\subsection{Scalability}

We imagine \app being used by a developer, or team of developers, in their everyday CI workflow.
For this reason the verification process must be fast. 
One of the benefits of using decision tree learning that that checking a new sample is a simple traversal of the tree, which takes negligible time.
We also would like the learning process to fast enough to scale to large datasets, however learning does not need to occur as frequently.
The learning process might be run overnight so that everyday the developer is using an updated model for the checking the CI configuration correctness.
To this end, we need to only ensure that the learning process is fast enough to run overnight.
We find that \app scales roughly linearly (dependant on the size and complexity of the repositories in the training set) and list training times in Table~\ref{table:training}.
All the experiments in this section are conducted on a MacBook Pro equipped with Haswell Quad Core i7-4870HQ 2.5 GHz CPU, 16GB memory, and PCIe-based 512 GB SSD harddrive.

Although our results with the repositories in Table~\ref{table:accuracy} already provide 83\% accuracy, it is generally the nature of machine learning to return better results with more data, which requires fast training times.
As a point of implementation, decreasing the constant factor cost of \app could improve running times so that we can scale to a truly large scale, and leverage the thousands of repositories with a total of more than 2.6 million commits available in the TravisTorrent dataset~\cite{travis-torrent}.
We leave this to future work, noting that this first version of \app focused entirely on correctness of implementation and not on any optimizations.

\begin{table}[tpb!]
\centering
\caption{Time for \app to learn a decsion tree model over various training set sizes.}
\label{table:training}
\setlength{\tabcolsep}{1em}
\begin{tabular}{|c|c|c|}
\hline
\multicolumn{2}{|c|}{\bf Training Set Size} & {\bf \app (sec)} \\
\# Repos & \# Commits & \\ 
\hline
\hline
1 & 10    & 11.094   \\ \hline
1 & 20   &  34.566  \\ \hline
1 & 30  &  50.070  \\ \hline
1 & 40  & 68.700   \\ \hline
1 & 50  &  85.740  \\ \hline
\hline
5 & 50  & 401.043   \\ \hline
10 & 50  & 1293.789  \\ \hline
15 & 50  &  1670.671 \\ \hline
20 & 50  &  2015.829  \\ \hline
25 & 50  &  2728.735  \\ \hline
\end{tabular}
\end{table}

\section{Related Work}

Configuration verification and validation 
has been considered as a promising way  
to tackling software failures resulting from
misconfiguration~\cite{xu15systems}.
Nevertheless, the strategy for generating models and checking
configuration settings still remains an open problem.

\para{Continuous Integration Build Prediction.}
The increasing prevalence of CI as a core software development tool 
has inspired significant work on the topic.
Some work has included predicting the status of a commit based on metadata such as the previous commit and history of the committer~\cite{ni2017cost,wolf2009predicting,hassan2006using}.
Natural language processing and sentiment analysis has also been used to predict build status~\cite{Paixao2017}.
However neither of these approaches provide the user with information 
that they are able to change that could actually fix the build. 
For example, if a user pushes a commit with a commit message that is 
``negative'' (\eg, ``annoying hack to get javascript working''), 
changing the commit message cannot change the build status.

Other work has predicted the time of a build, allowing developers to more effectively plan to work around long builds~\cite{bisong2017built}.
This is further evidence of the importance of providing developers with quick feedback on their CI builds.
While this work can help programmers more effectively manage their time, it cannot provide guidance for the user to actually make appropriate fixes.
Further work identified features of commits (such as complexity of the commit) that are statistically significantly correlated to build failure~\cite{islam2017insights}, but this work again does not attempt to help users fix these errors.
In contrast, \app specifically predicts the build status based on direct code features that users have control over, so that in addition to a predicted build status, users can to address the issues with the justification of the classification that \app provides.

A rich area for future work is to combine \app with the aforementioned techniques, so that we can gain stronger predictive results with metadata.
We may be able to use \app to then provide justifications for the classification of other tools, which rely on metadata that users are not able to control.

\para{Learning-based configuration verification.}
Several machine learning-based misconfiguration detection efforts 
also have been proposed~\cite{zhang14encore, SantolucitoCAV16, SantolucitoOOPSLA17, wang04automatic}.
For example, EnCore~\cite{zhang14encore} is the first learning-based
software misconfiguration checking approach.
The learning process is guided by a set of predefined rule templates
that enforce learning to focus on patterns of interest.
In this way, EnCore filters out irrelevant information and reduces
false positives; moreover, the templates are able to express
system environment information that other machine learning
techniques (\eg, \cite{wang04automatic}) 
cannot handle.
Later, ConfigC~\cite{SantolucitoCAV16} and 
ConfigV~\cite{SantolucitoOOPSLA17} are proposed to verify other 
types of configuration errors, including missing errors,
correlation errors, ordering errors, and type errors, 
in database systems.

Compared with the prior work, \app has the following 
differences and advantages.
First, \app targets configuration files for software building
process, which is a fundamentally different purpose from
the above three efforts. EnCore, ConfigC, and ConfigV are proposed
to verify configuration files for database system setup. 
Second, because the above three efforts only target the configuration
files for database systems, their model is specific to 
the key-value assignment representation, which is more structured
schema representation in the CI configurations.
Finally, the employed learning approaches in \app are quite different
from the three previous efforts.

\para{Language-support misconfiguration checking.}
There have been several language-based efforts  
proposed for specifying the correctness of
system-wide configurations. 
%introduced by fundamental deficiencies in
%either untyped or low-level languages. 
For example, in the datacenter network
management field, the network administrators often
produce configuration errors in their routing configuration files.
PRESTO~\cite{enck07configuration} 
automates the generation of device-native configurations
with configlets in a template language. 
Loo {\em et al.}~\cite{loo05declarative} adopt Datalog to reason about 
routing protocols in a declarative fashion. 
COOLAID~\cite{chen10declarative} constructs
a language abstraction
to describe domain knowledge about network devices and
services for convenient network management.
In software configuration checking area,
Huang {\em et al.}~\cite{huang15confvalley} proposed a specification 
language, ConfValley, for the administrators to write their
specifications, thus validating 
whether the given configuration files 
meet administrators' written specifications. 
Compared with the above efforts, \app focuses on 
the configuration files used for the software building process. 
In addition, another important purpose our work wants to achieve
is to automate configuration
verification process.
The above efforts can only offer language representations and 
still require the administrators to
manually write specifications, which is an error-prone
process. On the contrary, \app is able to automatically generate
specification rules.

\para{White-box based misconfiguration detection.}
White-box based misconfiguration detection techniques 
aim at checking configuration errors by analyzing the correlations
between the source code and configuration parameters.
The key intuition in white-box based detection 
is to emulate how a program uses configuration parameters,
thus understanding whether some constraints will be violated.
% Most existing detection approaches check 
%the configuration files against a set of predefined correctness 
%rules, named constraints, and then report errors if 
%the checked configuration parameters do not satisfy these rules.
For example, PCheck~\cite{xu16early} emulates potential commands 
and operations of the target system, and then adds configuration
checking or handling code to the system source code
in order to detect the errors before system fault.
This emulation is a white-box approach and 
requires access to the system's source code.
One drawback to the white-box based misconfiguration detection
approach is that for some systems (\eg, ZooKeeper) whose behavior is 
hard to emulate, PCheck cannot automatically generate 
the corresponding checking code.
Due to the emulation based testing strategy, PCheck's scope is 
limited to system reliability problems caused by 
misconfiguration parameters. 
In contrast, \app is a ``black-box'' approach 
and only requires a training set of configuration files to learn rules.
By using a rule learning strategy of examples, 
\app is able to detect general misconfiguration 
in building process .
%that are outside the scope of emulation.

\para{Misconfiguration diagnosis.}
Misconfiguration diagnosis approaches 
have been proposed to address configuration problems post-mortem.
For example, ConfAid~\cite{attariyan10automating} 
and X-ray~\cite{attariyan12x-ray} use dynamic information
flow tracking to find possible configuration errors 
that may have resulted in
failures or performance problems. AutoBash~\cite{su07autobash} 
tracks causality and automatically fixes 
misconfigurations. Unlike \app, most misconfiguration
diagnosis efforts aim at finding errors after system
failures occur, which leads to prolonged lost server time.

\section{Conclusions}

This paper has presented a practical tool \app that can 
automatically check the errors
in CI configuration files before the build process.
Driven by the insight that repositories in CI environment have
lists of build status histories,
our approach automatically generates specification for the CI configuration
error checking by learning the repository histories. 
We evaluate \app on real-world data from GitHub and find that
we have 83\% accuracy of predicting a build failure.

\para{The potential impact and future work.}
In recent years, Continuous Integration (CI) testing has changed the workflow for developers in testing their code before the code is integrated into a company-wide codebase.
It is, therefore, very important to understand how a new tool, such as \app, could be integrated into the CI workflow and the impact it would have on developer workflow.
Most immediately, \app can predicate failures and provide developers with a location and reason for the expected failure, so they may fix the error.
This can reduce the amount of time that developers need to wait for the CI build to complete, which can take hours.
\app could also be combined with post-failure analysis strategies, such as log analysis, to help developers more quickly find their error.

In order to understand this impact better, we are planning to deploy \app over a long term study in large scale of CI environments.
As \app runs as a client side static analysis tool, it is important to see the impact \app has for developers in the moment of pushing code.
In particular, we plan to investigate the cost of false positive reports, and the best way to present our error messages to developers.

\com{

%% Acknowledgments
\begin{acks}                            %% acks environment is optional
                                        %% contents suppressed with 'anonymous'
  %% Commands \grantsponsor{<sponsorID>}{<name>}{<url>} and
  %% \grantnum[<url>]{<sponsorID>}{<number>} should be used to
  %% acknowledge financial support and will be used by metadata
  %% extraction tools.
  This material is based upon work supported by the
  \grantsponsor{GS100000001}{National Science
    Foundation}{http://dx.doi.org/10.13039/100000001} under Grant
  No.~\grantnum{GS100000001}{nnnnnnn} and Grant
  No.~\grantnum{GS100000001}{mmmmmmm}.  Any opinions, findings, and
  conclusions or recommendations expressed in this material are those
  of the author and do not necessarily reflect the views of the
  National Science Foundation.
\end{acks}

}

%% Bibliography
\bibliography{travis,os}

%\newpage

%% Appendix
%\appendix
%\section{Appendix}

\end{document}